# Vertically Coupled Double Quantum Dots Connected In Parallel

Shinichi Amaha[1], Tsuyoshi Hatano[2], Takashi Nakajima[1], and Seigo Tarucha[1]

[1] *RIKEN Center for Emergent Matter Science (CEMS), Wako, Saitama 351-0198, Japan*
[2] *College of Engineering, Nihon University, 1 Nakagawara, Tokusada, Tamura, Koriyama, Fukushima, 963-8642, Japan*

We report charge transport measurements in a ring-shaped quadruple quantum dot system, composed of two vertically coupled double quantum dots connected in parallel. The vertical coupling introduces an isospin degree of freedom tied to the layer index, and the parallel configuration enables independent access to each quantum dot pair. This design allows us to observe Coulomb diamonds and evaluate the interlayer energy offset. By extending this platform to triangular and hexagonal artificial lattices, we explore correlation effects such as isospin frustration. These results highlight the system's potential for studying interaction-driven quantum phases.

## 1. Introduction

Quantum dots (QDs) are regarded as artificial atoms due to their electronic properties associated with three-dimensional electron confinement [1, 2]. Owing to their discrete energy levels and high tunability, QDs have emerged as a promising platform for quantum information processing [2-5] and the simulation of strongly correlated electron systems [6-8]. In particular, multiple QD arrays allow for control of geometrical configurations and inter-dot tunnel couplings, enabling the experimental realization of theoretical models such as the Hubbard model [7-9]. This has facilitated the exploration of various quantum phases driven by electron correlations.

One representative example of such correlation-driven quantum phases is geometric frustration, a phenomenon in which the spatial arrangement of dots prevents simultaneous minimization of all interaction energies, leading to degenerate ground states and nontrivial quantum effects. In triangular triple QDs, experimental signatures believed to originate from geometric frustration—such as isotropic charge transport and degenerate charge configurations in stability diagrams—have been reported [10].

Another example of correlation-induced phenomena is Nagaoka ferromagnetism [7], which arises in the Hubbard model with a single hole under strong on-site Coulomb repulsion, leading to spontaneous ferromagnetic ordering. QD systems are particularly well-suited to testing this scenario due to their ability to finely control electron numbers and tunnel couplings. Indeed, recent experimental work has demonstrated spontaneous ferromagnetic spin alignment in a four-site QD plaquette with one hole, providing Nagaoka ferromagnetism in an artificial lattice [7].

In this study, we report an experimental investigation of a ring of four QDs, realized by two vertically coupled double QDs (DQDs) connected in parallel. Increasing the number of vertically stacked layers [11, 12] enhances the system's degrees of freedom and enables exploration of richer quantum phenomena associated with interlayer coupling and symmetry breaking. For example, in two-layer systems, the layer index introduces a binary quantum degree of freedom, often referred to as isospin [13, 14]. The presence of this isospin degree of freedom enables the exploration of rich quantum behaviors that arise from its coupling with real spin and charge dynamics. We further extend our investigation to six-dot systems, including three vertically coupled DQDs arranged in a triangular lattice and artificial graphene structures composed of DQDs forming a hexagonal lattice, aiming to deepen our understanding of electronic behavior in systems with multiple, interacting quantum degrees of freedom.

## 2. Transport Characteristics of Vertical DQDs coupled in Parallel

*2.1 Device Structure*

Figure 1(a) shows a schematic diagram of a quadruple quantum dot (QQD) device consisting of two vertically coupled double quantum dots (DQDs) connected in parallel. The device comprises two pillar-shaped mesas, each incorporating two quantum wells separated by a thin tunneling barrier, thereby forming a vertically coupled DQD.

Each mesa is equipped with four surface Schottky gates—$G_L$, $G_R$, $G_{C1}$, and $G_{C2}$—as shown in Fig. 1(a). Note that $G_{C2}$ is located on the backside of the illustrated perspective and is therefore not visible in Fig. 1(a). The $G_L$ and $G_R$ gates are primarily used, by applying gate voltages $V_{GL}$ and $V_{GR}$, respectively, to control the electro-chemical potentials in the corresponding mesas, while each mesa contains a vertically coupled DQD (DQD-L, DQD-R). $G_{C1}$ and $G_{C2}$ are used to tune the tunnel coupling between the two mesas. This gate configuration follows the design commonly used in conventional parallel-coupled vertical DQD devices [9], which are typically fabricated using a single quantum well structure. Unlike conventional devices based on a single quantum well [9], the present structure enables vertical DQD formation within each mesa due to the inclusion of two quantum wells.

Figure 1(b) shows a schematic diagram of the current flow in the QQD device. As illustrated, each of the two pillar-shaped mesas is connected to both the source and drain electrodes. A bias voltage $V_d$ is applied between the source and drain, allowing current to flow vertically through each mesa, as indicated by black solid arrows in Fig. 1(b). However, due to the presence of inter-dot tunneling between the QDs formed in the two mesas, the current paths are not strictly confined within each mesa. Electrons may tunnel between dots in different mesas, as indicated by gray solid arrows in Fig. 1(b), resulting in more complex transport behavior than that of two independent vertical channels.

We adopted the material with the inter-layer barrier thickness equals 5.5nm where the vertical coupling strength is roughly estimated about $2t \sim 0.4$meV [14, 15]. In contrast, the lateral coupling strength in conventional parallel-coupled vertical DQD devices is typically about $2t \sim 0.2$meV [9]. Therefore, in our device, the interlayer coupling is stronger than the lateral coupling, indicating that quantum mechanical interactions predominantly occur along the vertical direction. The gate voltages $V_{GL}$, $V_{GR}$, and $V_{GC}$ applied to the gate electrodes $G_L$, $G_R$, and $G_C$, independently, and the current $I_d$ through the QQD device is measured under the temperature $T \sim 1.5$K.

## 2.1 Transport Properties

### 2.1.1 Stability diagrams of QQD tuned by center gate voltages

Figure 2 (a) displays a charge stability diagram, where the current $I_d$ is mapped as a function of $V_{GL}$ and $V_{GR}$ under $V_{sd} \sim 150\mu$eV and $V_{GC}$=-0.8V. The color scale represents the magnitude of the current $I_d$, with blue indicating zero current and progressing through white and red to black as the current increases. Regions of zero current ($I_d \sim 0$) correspond to Coulomb blockade, indicating stable electron configurations in the QQDs. These blockade regions form a honeycomb-like pattern similar to that observed in conventional DQD systems. At the boundaries between Coulomb blockade regions, where current is expected to flow, anti-crossing features are observed (See ▲ in Fig. 2(a)), similar to those in conventional DQD systems [16]. A notable difference from conventional parallel-coupled DQD systems is that, in the QQD stability diagram, the observed current $I_d$ is significantly reduced or even completely vanishes at specific sections of the Coulomb blockade boundary, where a sufficiently strong current would normally be expected in case of conventional parallel-coupled DQD (See [9]).

Figure 2(b) displays a charge stability diagram, where the current $I_d$ is mapped as a function of the gate voltages $V_{GL}$ and $V_{GR}$, under a source-drain bias of $V_{sd} \sim 150\mu$eV and a center gate voltage of $V_{GC}$=−1.0V. Similar to Fig. 2(a), characteristic Coulomb blockade regions are observed, forming a honeycomb-like pattern that reflects the discrete charge states of the coupled quantum dot system. However, compared to the case in Fig. 2(a), the anti-crossing features observed at the boundaries between blockade regions (see ▲ in Fig. 2(b)) appear less pronounced. This reduction in curvature is attributed to the decreased tunnel coupling between the quantum dots located in different mesas, as a result of the lower $V_{GC}$ voltage.

2.1.2 *Assignment of the offset between vertically coupled QDs*

To investigate the origin of the current reduction observed at certain Coulomb blockade boundaries, we performed Coulomb diamond measurements by sweeping gate voltages along the A–B and C–D lines indicated in Fig. 2 (a). The A–B line corresponds to a gate voltage trajectory where the electron number in one of the mesas (i.e., one of the DQDs) is expected to be zero, while the C–D line corresponds to the case for the other mesa as shown in Fig. 2 (a). These directions allow for independent access to the transport characteristics of each DQD, because one DQD is effectively depleted, and the electron number in the remaining DQD is predominantly controlled by the corresponding gate voltage (i.e. $V_{GR}$ for DQD-R and $V_{GL}$ for DQD-L).

Figure 3 (a) shows the differential conductance $dI_d/dV_{sd}$ measured as a function of the source-drain bias $V_{sd}$ and gate voltage $V_{GR}$ along the A–B line. The obtained stability diagram exhibits a series of distorted Coulomb diamonds, consistent with those typically observed in serially coupled DQD systems. Notably, one vertical line (■) and one kink (●) are observed at the boundaries of the diamonds, which reflect the existence of offset energy between the two vertically coupled quantum dots [16].

In order to evaluate the energy offset between the two QDs in DQD-R, we employed the constant interaction (CI) model. Figure 3(b) shows the calculated Coulomb blockade region (indicated in white), in which a kink (●) and a vertical line (■) are clearly reproduced. These features reflect the characteristic charge stability pattern in the presence of a finite energy offset. Based on this analysis, the energy offset $\Delta_R$ between the two QDs in DQD-R can be estimated as $\Delta_R \sim U/2$. $U$ denotes the intradot Coulomb charging energy [17].

The charge configurations in regions α and β of the observed Coulomb diamonds are assigned, based on the CI model, as $(N_{RU}, N_{RD})$ = (0, 0) for region α and (0, 1) for region β, where $N_{RU}$ and $N_{RD}$ represent the electron numbers in the upper and lower QDs of DQD-R, respectively.

It is also noted that the electron number in DQD-L is expected to be zero in the region to the right of the dotted line. Therefore, the overall charge states of the QQD can be assigned as $(N_{RU}, N_{RD}, N_{LU}, N_{LD})$ = (0, 0, 0, 0) in region α and (0, 1, 0, 0) in region β, respectively.

Figure 3(c) also shows the differential conductance $dI_d/dV_{sd}$ measured as a function of the $V_{sd}$ and $V_{GL}$ along the C–D line. Similar to Fig. 3(a), the resulting stability diagram exhibits a series of distorted Coulomb diamonds characteristic of serially coupled DQD systems. In this case, three kinks (○$_1$, ○$_2$ and ○$_3$) are observed, indicating the presence of an energy offset $\Delta_L$ between the upper and lower QDs in DQD-L. Following the same analysis based on the CI model as shown in Fig. 3(d), $\Delta_L$ is estimated to be approximately $\Delta_L \sim 3U/2$.

Comparing Fig. 3(c) to Fig. 3(d), the corresponding charge states in regions γ, δ, ε and ζ in Fig. 3(d) are assigned as ($N_{LU}$, $N_{LD}$) = (0, 0), (0, 1), (0, 2) and (0, 3) respectively. Since the electron number in DQD-R is expected to be zero along the C–D line, the total charge configurations of the QQD system can be assigned as ($N_{RU}$, $N_{RD}$, $N_{LU}$, $N_{LD}$) = (0, 0, 0, 0) in region γ, (0, 0, 0, 1) in region δ, (0, 0, 0, 2) in region ε and (0, 0, 0, 3) in region ζ.

The Coulomb diamond measurement under zero electron regime with one of DQD in our QQD provides the bias spectroscopy to detect the offset between QDs in one of DQD. The bias spectroscopy allows the assignment of charge configurations under the Coulomb blockade region in the charge stability diagram obtained by sweeping the gate voltages $V_{GL}$ and $V_{GR}$ as shown in Fig. 2(a). The resulting charge states, summarized schematically in Fig. 4, indicate the presence of electrons in each of the QD in our QQD. Note that a small current $I_d$ is observed in Fig. 2(a), even though the upper QD in DQD-L is nominally empty and direct transport is blocked. This small current $I_d$ can be attributed to cotunneling processes that bypass the empty dot, enabling weak current flow through the DQD [12].

Charge-state identification revealed the configurations ($N_{RU}$, $N_{RD}$, $N_{LU}$, $N_{LD}$) = (1, 1, 0, 0) and (0, 1, 0, 1). In a serial DQD, the charge states ($N_1$, $N_2$) = (1,1) and (0,2) (or (2,0)) are adjacent in the characteristic honeycomb diagram ($N_1$ and $N_2$ denote electron numbers of each QD), and Pauli spin blockade (p-SB) is typically observed at the boundary between these states. However, in Fig. 4, the charge states (1,1,0,0) and (0,1,0,1)—rather than (0,2,0,0)—are adjacent. In this case, p-SB is unlikely to occur because a competing transition from (1,1,0,0) to (0,1,0,1), which involves both interlayer and inter-mesa tunneling, allows the triplet state to remain accessible for transport.

The observation of the lateral coupling between DQD-L and DQD-R assigned in Fig. 1 (a) and (b) indicate that all four QDs are coupled in a ring-like configuration via vertical and lateral tunnel and capacitive couplings. And, as shown in Fig. 4, the charge configuration (1,1,0,1), was also experimentally identified, corresponding to a three-electron occupation in the four-dot plaquette. While this filling satisfies the condition for realizing Nagaoka ferromagnetism in an ideal system with uniform site potentials, the presence of finite interlayer energy offsets $\Delta_L$ and $\Delta_R$ in the present device breaks the site equivalence required to stabilize a fully spin-polarized ground state. These energy offsets suppress the coherent delocalization of the single hole, which is essential for gaining kinetic energy in the Nagaoka mechanism, and thereby might prevent the realization of the Nagaoka state under the current device conditions.

## 3. Multi-Dot Lattices Formed from Vertically Coupled DQDs

*3.1 Triangular Array of Vertically Coupled DQDs*

Building upon the demonstrated hybrid configuration combining vertical and lateral couplings, we now extend our investigation to more complex QD lattices formed by vertically coupled DQDs. Such structures serve as a promising platform for simulating quantum many-body systems, where the interplay between spatial geometry and interlayer coupling can give rise to rich quantum phenomena.

One of the QD lattices formed by vertically coupled DQDs is a triangular array of three vertically coupled DQDs (DQD-1, DQD-2, DQD-3) as schematically shown in Figure 5 (a). Note that a triangular array of three vertical QDs is performed experimentally [18], so that triangular array of three vertically coupled DQDs could be realized. Each site hosts a single electron under strong on-site Coulomb repulsion, effectively freezing charge fluctuations. In this regime, spin degrees of freedom can be neglected, and the relevant low-energy dynamics are governed by the layer index, described as an isospin with $I_z = +1/2$ (bonding state) and $I_z = -1/2$ (antibonding state).

Assuming the interlayer tunnel coupling *t* is enough small compared to the inter-dot Coulomb energy *U* and intra-dot Coulomb energy *V*, we apply second-order perturbation theory to derive an effective interaction between neighboring isospins. An interlayer energy offset Δ is also included, acting as a longitudinal field on the isospin.

The resulting effective Hamiltonian is given by:

$$H_{eff} = J_{eff} \sum_{<i,j>} I_i^z I_j^z + \Delta \sum_i I_i^z$$

, where $J_{eff} \sim 4t^2/(U-V)$.

The triangular geometry and the antiferromagnetic nature of $J_{eff}$ give rise to geometric frustration in the isospin degrees of freedom '*Ising isospin fluctuation*', potentially leading to a degenerate ground state and rich many-body phenomena. Importantly, the energy offset Δ between the upper and lower layers plays a critical role in determining whether such frustration is realized. A large $\Delta$ tends to polarize the isospin toward one layer, suppressing the superposition of bonding and antibonding states and thereby lifting the degeneracy. Conversely, when $\Delta$ is sufficiently small compared to $J_{\text{eff}}$, the system retains isospin degeneracy and can support frustration-induced quantum phases. Thus, experimental determination of $\Delta$, as demonstrated in our parallel-coupled double QD system, provides essential information for assessing the feasibility of realizing geometrically frustrated isospin configurations.

*3.2 Hexagonal Array of Vertically Coupled DQDs*

It is conceivable to fabricate a hexagonal lattice device, as schematically illustrated in Figure 5 (b), by arranging vertically coupled DQDs in a honeycomb geometry on a substrate equipped with double quantum wells. This artificial graphene structure mimics the lattice configuration of natural graphene, with the vertically stacked DQDs forming the sites of a honeycomb lattice.

In graphene, carbon atoms naturally form a two-dimensional honeycomb lattice, whose electronic structure exhibits linear energy dispersion near the Dirac points—resulting in massless Dirac fermions. Artificial graphene [19], in contrast, consists of lithographically defined QDs, whose site energies, inter-dot tunnel couplings, and interlayer energy offsets can be independently tuned via gate voltages, allowing for highly flexible control over system parameters.

A key structural difference between the two systems lies in the vertical alignment between layers. In natural graphene, the atomic positions of adjacent layers cannot be perfectly aligned due to lattice symmetry and energetic constraints. This leads to typical stacking sequences such as ABA (Bernal) or ABC (rhombohedral), where atoms in one layer are laterally displaced relative to those in the neighboring layers. These displacements modulate the interlayer tunnel couplings and influence the overall band structure.

In contrast, artificial graphene enables precise vertical alignment of QD positions between layers using nanofabrication and epitaxial techniques. This capability allows the formation of a layered QD lattice in which each upper dot is directly above its lower counterpart—something that cannot be realized in natural graphene. If an individual electrode can be attached to one of the pillars, it becomes possible to realize tunable "adatom-like" features in artificial graphene lattices, analogous to placing adatoms on natural graphene.

Such vertical alignment permits well-defined interlayer tunneling and the formation of bonding and antibonding molecular states, leading to the emergence of a controllable degree of freedom known as isospin. The ability to engineer and manipulate these vertically coupled systems offers new opportunities to explore coherent quantum effects and interlayer interactions, demonstrating a fundamental advantage of artificial graphene over its natural counterpart.

## 4. Conclusion

We have studied charge transport in a four-quantum-dot ring structure composed of two vertically coupled double quantum dots connected in parallel. The device structure enabled the extraction of the interlayer energy offset, which arises from asymmetries in confinement or electrostatic potential. Coulomb diamond measurements allowed us to identify characteristic features such as vertical lines and kinks, providing a clear signature of offset-induced charge configurations. These results confirm that the interlayer energy asymmetry plays a critical role in determining the local charge distribution and accessible spin states.

We also discussed model systems based on triangular and hexagonal arrays of vertically coupled dots, and derived an effective isospin Hamiltonian for the triangular case. The results suggest that interlayer coupling and energy offset play a key role in determining the ground state configuration and quantum correlations. These findings indicate that systems composed of vertically coupled quantum dots connected in parallel provide a versatile platform for investigating emergent correlated electron phenomena in lattice geometries.

**Acknowledgment**

We are grateful to T. Otsuka, M. Delbecq, J. Yoneda, K. Takeda, and A. Noiri for helpful discussions and insightful suggestions. We thank D. G. Austing for providing the material. T. H. was supported by Nihon University Multidisciplinary Research Grant for (2021).

Figure 1.

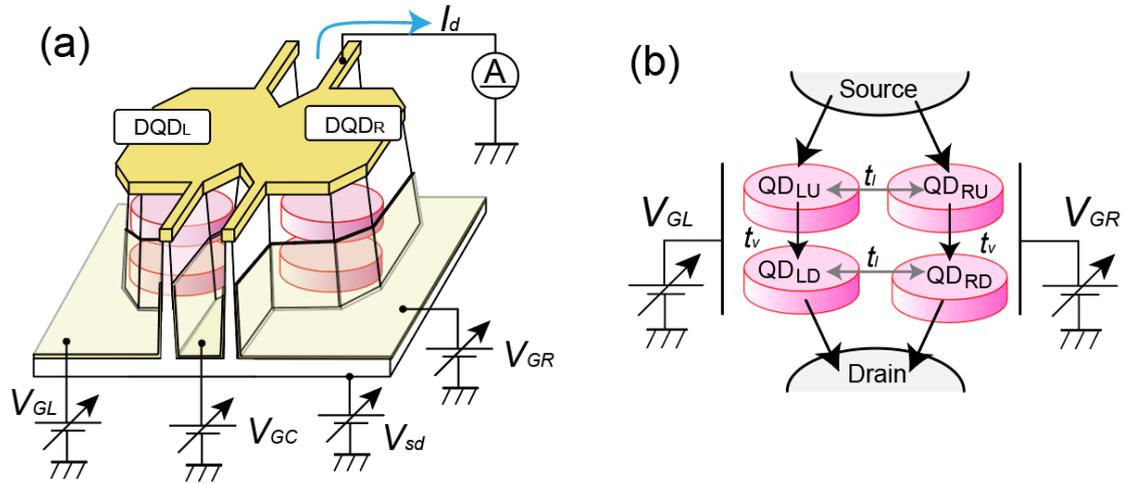

(a) Schematic diagrams of a QQD device consisting of DQD-L and DQD-R connected in parallel. Only one of the two center gates $V_{GC}$ is shown; the other is located on the back side and not depicted. (b) The current flow in the QQD device. Note that two center gate electrodes are attached to the

Figure 2.

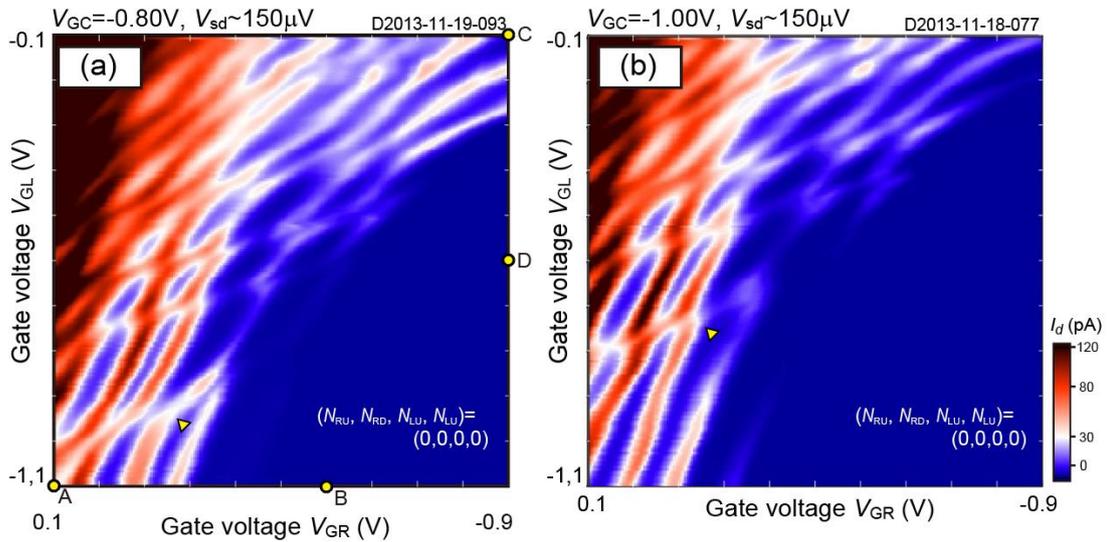

Current $I_d$ is mapped as a function of $V_{GL}$ and $V_{GR}$ under $V_{sd} \sim 150\mu eV$ with (a) $V_{GC}$=-0.8V and (b) $V_{GC}$=-1.0V.

Figure 3.

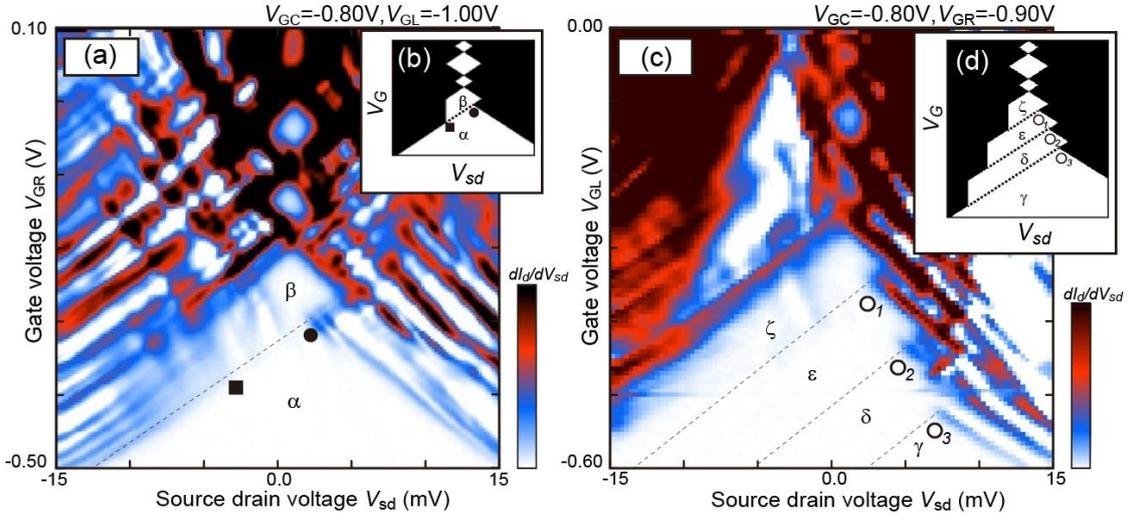

(a) Differential conductance $dI_d/dV_{sd}$ as a function of source-drain voltage $V_{sd}$ measured along the A–B line in Fig. 2 (a). (b) Corresponding Coulomb diamonds for (a), calculated using the constant interaction (CI) model. (c) Differential conductance $dI_d/dV_{sd}$ as a function of $V_{sd}$ measured along the C–D line in Fig. 2 (a). (d) Corresponding Coulomb diamonds for (c), based on the CI model.

Figure 4.

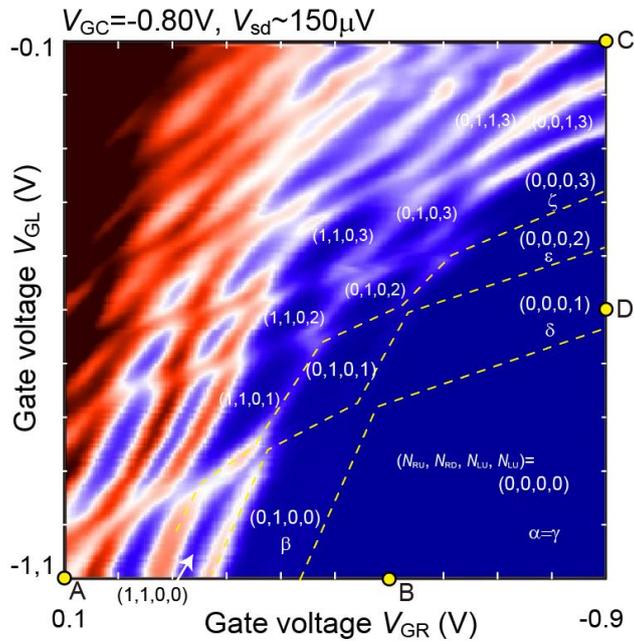

Summarized charge configuration ($N_{RU}$, $N_{RD}$, $N_{LU}$, $N_{LD}$) in Figure 2. (a), where $N_{RU}$, $N_{RD}$, $N_{LU}$ and $N_{LD}$ mean the electron number of quantum dot RU, RD, LU and LD, respectively.

Figure 5.

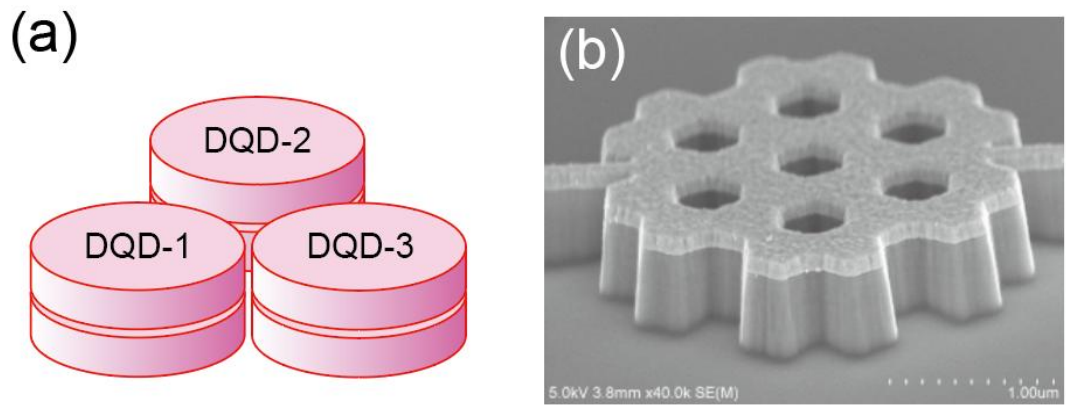

(a) Schematic diagram of triangular array of three vertically coupled DQDs. (b) SEM image of an artificial graphene sample composed of vertically QD arranged in a honeycomb geometry without gates.